\documentclass[twocolumn]{autart}    

\usepackage{graphicx}

\usepackage[sort,compress]{cite}

\usepackage{latexsym}
\usepackage{psfrag}
\usepackage{amssymb,amscd,amssymb,amsmath,mathrsfs}
\usepackage{overpic}
\usepackage{mdwlist}           
\usepackage{ifpdf}
\usepackage{caption2} 
\usepackage{url} 
\usepackage{float}
\usepackage{color}

\makeatletter
\newcommand{\rmnum}[1]{\romannumeral #1}
\newcommand{\Rmnum}[1]{\expandafter\@slowromancap\romannumeral #1@}
\makeatother

\begin{document}

\begin{frontmatter}

\title{Supervision Localization of Timed Discrete-Event Systems}

\thanks[footnoteinfo]{This work was supported
in part by the State Key Laboratory of Electrical Insulation and
Power Equipment (China), and by the Natural Sciences and Engineering
Research Council (Canada), Grant no. 7399.}

\author[XJTU]{Renyuan Zhang}\ead{r.yuan.zhang@gmail.com},    
\author[UofT]{Kai Cai}\ead{kai.cai@scg.utoronto.ca},               
\author[XJTU]{Yongmei Gan}\ead{ymgan@mail.xjtu.edu.cn},  
\author[XJTU]{Zhaoan Wang}\ead{zawang@mail.xjtu.edu.cn},
\author[UofT]{W.M. Wonham}\ead{wonham@control.utoronto.ca}

\address[XJTU]{School of Electrical Engineering, Xi'an Jiaotong University, Xi'an, Shaanxi 710049, China}  
\address[UofT]{Department of Electrical and Computer Engineering, 
University of Toronto, Toronto, ON M5S 3G4, Canada}    



\begin{abstract}                          
We study supervisor localization for real-time discrete-event
systems (DES) in the Brandin-Wonham framework of timed supervisory
control. We view a real-time DES as comprised of asynchronous agents
which are coupled through imposed logical and temporal
specifications; the essence of supervisor localization is the
decomposition of monolithic (global) control action into local
control strategies for these individual agents. This study extends
our previous work on supervisor localization for untimed DES, in
that monolithic timed control action typically includes not only
disabling action as in the untimed case, but also ``clock
preempting'' action which enforces prescribed temporal behavior. The
latter action is executed by a class of special events, called
``forcible'' events; accordingly, we localize monolithic preemptive
action with respect to these events. We demonstrate the new features
of timed supervisor localization with a manufacturing cell case
study, and discuss a distributed control implementation.
\end{abstract}

\end{frontmatter}

\section{Introduction} \label{sec:intro}

Recently we developed a top-down approach, called \emph{supervisor
localization} \cite{CaiWonham:2010a,CaiWonham:2010b} to the
distributed control of untimed discrete-event systems (DES) in the
Ramadge-Wonham (RW) supervisory control framework
\cite{RamadgeWonham:87,Wonham:2011a}. We view the plant to be
controlled as comprised of independent asynchronous agents which are
coupled implicitly through logical control specifications. To make
the agents smart and semi-autonomous, our localization algorithm
allocates \emph{external} supervisory control action to individual
agents as their \emph{internal} control strategies, while preserving
the optimality (maximal permissiveness) and nonblocking properties
of the overall monolithic (global) controlled behavior. Under the
localization scheme, each agent controls only its own events,
although it may very well need to observe events originating in
other (typically neighboring) agents.

In this paper we extend the supervisor localization theory to a
class of \emph{real-time} DES, and address distributed control
problems therein. Many time-critical applications can be modeled as
real-time DES, such as communication channels, sensor networks,
scheduling and resource management \cite{LeeLeuSon:07}. Typical
timing features include communication delays and operational hard
deadlines. The correctness and optimality of real-time DES depend
not only on the system's logical behavior, but also on the times at
which various actions are executed. Moreover, rapid advances in
embedded, mobile computation and communication technologies
\cite[Part~III]{LeeLeuSon:07} have enabled distributed
implementation of control algorithms.  These developments jointly
motivate this study of supervisor localization for real-time DES.

A variety of real-time DES models and approaches are available.
Notable works include Brave and Heymann's ``clock automata''
\cite{BraveHeymann:88}, Ostroff's ``timed transition models''
\cite{Ost:90}, Brandin and Wonham's timed DES (TDES)
\cite{BrandinWonham:94}, Cassandras's ``timed state automata''
\cite{Cassandras:93}, Wong-Toi and Hoffman's model based on ``timed
automata'' \cite{WongToiHoffman:91}, and Cofer and Garg's model
based on ``timed Petri nets'' \cite{CoferGarg:96}. We adopt Brandin
and Wonham's TDES (or BW model) as the framework for developing a
timed supervisor localization theory for two reasons. First, the BW
model is a direct extension from the RW framework (where our untimed
localization theory is based), retaining the central concepts of
controllability, and maximally permissive nonblocking supervision.
This feature facilitates developing a timed counterpart of
supervisor localization. Second, the BW model captures a variety of
timing issues in a useful range of real-time discrete-event control
problems \cite{BrandinWonham:94},\cite[Chapter~9]{Wonham:2011a}.
While it may be possible to develop supervisor localization in an
alternative framework, as a preliminary step into real-time
supervisor localization we choose the BW model for its close
relation with previous work.

The principal contribution of this paper is the development of a
timed supervisor localization theory in the BW TDES framework, which
extends the untimed counterpart in
\cite{CaiWonham:2010a,CaiWonham:2010b}.  In this timed localization,
a novel feature is ``event forcing'' as means of control, in
addition to the usual ``event disabling''. Specifically,
``forcible'' events are present in the BW model as events that can
be relied on, when subject to some temporal specification, to
``preempt the tick of the clock'', as explained further in
Section~\ref{sec:2}. Correspondingly, in localizing the monolithic
supervisor's control action, we localize not only its disabling
action as in the untimed case, but also its preemptive action with
respect to individual forcible events.  Central to the
latter are several new ideas: ``local preemptor'', ``preemption
consistency relation'', and ``preemption cover''. We will prove
that localized disabling and preemptive behaviors collectively
achieve the same global optimal and nonblocking controlled behavior
as the monolithic supervisor does.  The proof relies on
the new preemption concepts and also controllability for TDES.
Moreover, the derived local controllers typically have much smaller
state size than the monolithic supervisor, and hence their disabling
and preemptive logics are often more transparent. We demonstrate
this empirical result by a case study of a manufacturing cell taken
from \cite{BrandinWonham:94}.

The paper is organized as follows. Section~\ref{sec:2} provides a
review of the BW TDES framework. Section~\ref{sec:3}
formulates the timed supervisor localization problem, and
Section~\ref{sec:4} presents a constructive solution procedure.
Section~\ref{sec:5:exmp} studies a manufacturing cell example; and finally,
Section~\ref{sec:conclusion} draws conclusions.

\section{Preliminaries on Timed Discrete-Event Systems} \label{sec:2}

This section reviews the TDES model proposed by Brandin and Wonham
\cite{BrandinWonham:94},\cite[Chapter~9]{Wonham:2011a}. First
consider the untimed DES model
\begin{equation} \label{e1}
{\bf G}_{act} = (A, \Sigma_{act}, \delta_{act}, a_0, A_m).
\end{equation}
Here $A$ is the finite set of {\it activities}, $\Sigma_{act}$ is
the finite set of {\it events}, $\delta_{act}:A \times \Sigma_{act}
\to A$ is the (partial) {\it activity transition function}, $a_0 \in
A$ is the {\it initial activity}, and $A_m \subseteq A$ is the set
of {\it marker activities}. Let $\mathbb{N}$ denote the set of
natural numbers $\{0,1,2,...\}$. We introduce \emph{time} into ${\bf
G}_{act}$ by assigning to each event $\sigma \in \Sigma_{act}$ a
{\it lower time bound} $l_\sigma \in \mathbb{N}$ and an {\it upper
time bound} $u_\sigma \in \mathbb{N} \cup\{\infty\}$, such that
$l_\sigma \leq u_\sigma$; typically, $l_\sigma$ represents a delay
in communication or in control enforcement, while $u_\sigma$ is
often a hard deadline imposed by legal specification or physical
necessity. With these assigned time bounds, the event set
$\Sigma_{act}$ is partitioned into two subsets: $\Sigma_{act} =
\Sigma_{spe} \dot\cup \Sigma_{rem}$ ($\dot\cup$ denotes
\emph{disjoint union}) with $\Sigma_{spe} := \{\sigma \in
\Sigma_{act}|u_\sigma \in \mathbb{N}\}$ and $\Sigma_{rem} :=
\{\sigma \in \Sigma_{act} | u_\sigma = \infty\}$; here ``spe''
denotes ``prospective'', i.e. $\sigma$ will occur within some
prospective time (with a finite upper bound), while ``rem'' denotes
``remote'', i.e. $\sigma$ will occur at some \emph{indefinite} time
(with no upper bound), or possibly will never occur at all.

A distinguished event, written $tick$, is introduced which
represents ``tick of the global clock''. Attach to each event
$\sigma \in \Sigma_{act}$ a (countdown) {\it timer} $t_\sigma \in
\mathbb{N}$, whose default value $t_{\sigma0}$ is set to be
\begin{equation} \label{e2}
t_{\sigma 0} := \left\{
             \begin{array}{lcl}
             u_\sigma &\text{if} &\sigma \in \Sigma_{spe}, \\
             l_\sigma &\text{if} &\sigma \in \Sigma_{rem}.
             \end{array}
        \right.
\end{equation}
When timer $t_\sigma>0$, it decreases by 1 (counting down) if event $tick$
occurs; and when $t_\sigma = 0$, event $\sigma$ must occur (resp.
may occur) if $\sigma \in \Sigma_{spe}$ (resp. if $\sigma \in
\Sigma_{rem}$). Note that while $tick$ is a global event, each timer
$t_\sigma$ is local (with respect to the event $\sigma$). Also
define the {\it timer interval} $T_\sigma$ by
\begin{equation} \label{e3}
T_\sigma := \left\{
             \begin{array}{lcl}
             {[0, u_\sigma]} &\text{if} &\sigma \in \Sigma_{spe}, \\
             {[0, l_\sigma]} &\text{if} &\sigma \in \Sigma_{rem}.
             \end{array}
        \right.
\end {equation}
Thus $t_\sigma \in T_\sigma$.

Based on (\ref{e1})-(\ref{e3}), the TDES model ${\bf G}$ is given by
\begin{equation} \label{e4}
{\bf G} := (Q, \Sigma, \delta, q_0, Q_m),
\end{equation}
where $Q := A \times \prod\{T_\sigma|\sigma \in \Sigma_{act}\}$
($\prod$ denotes \emph{Cartesian product}) is the finite set of
\emph{states}\footnote{An upper bound for the state size $|Q|$ is
$|A| * \prod\limits_{\sigma \in \Sigma_{act}} t_{\sigma0}$ (here
$\prod$ denotes \emph{scalar multiplication}), which in practice can
be much larger than its untimed counterpart $|A|$.}, a state $q \in
Q$ being of the form $q = (a, \{t_\sigma|\sigma \in \Sigma_{act}\})$
(i.e. a ($1+|\Sigma_{act}|$)-tuple); $\Sigma := \Sigma_{act}
\dot\cup \{tick\}$ is the finite set of events;
$\delta:Q\times\Sigma \rightarrow Q$ is the (partial) {\it state
transition function}; $q_0 = (a_0, \{t_{\sigma 0}|\sigma \in
\Sigma_{act}\})$ ($t_{\sigma 0}$ as in (\ref{e2})) is the {\it
initial state}; and $Q_m \subseteq A_m \times \prod\{T_\sigma|\sigma
\in \Sigma_{act}\}$ is the set of {\it marker states}.  Starting
from $q_0$, TDES ${\bf G}$ executes state transitions in accordance
with its transition function $\delta$.  Let $q=(a,\{t_\alpha|\alpha
\in \Sigma_{act}\}) \in Q$ and $\sigma \in \Sigma_{act}$; $\delta$
is defined at $(q,\sigma)$, written $\delta(q,\sigma)!$, if
$\delta_{act}$ of ${\bf G}_{act}$ is defined at $(a,\sigma)$ (i.e.
$\delta_{act}(a,\sigma)!$) and timer $t_\sigma$ satisfies (i) $0
\leq t_\sigma \leq u_\sigma-l_\sigma$ when $\sigma \in
\Sigma_{spe}$, and (ii) $t_\sigma=0$ when $\sigma \in \Sigma_{rem}$.
The new state $q'=\delta(q,\sigma)$ is given by
$q'=(\delta_{act}(a,\sigma), \{t'_\alpha|\alpha \in
\Sigma_{act}\})$, where $t'_\sigma$ is set to be its default value
$t_{\sigma 0}$ as in (\ref{e2}); for other timers $t_{\alpha}$,
$\alpha \neq \sigma$, the reader is referred to detailed updating
rules given in \cite{BrandinWonham:94,Wonham:2011a}. On the other
hand, $\delta(q, tick)!$ if no timer of a prospective event is zero,
and $q'=\delta(q, tick)=(a,\{t'_\alpha|\alpha \in \Sigma_{act}\})$,
i.e. there is no change in the activity component $a$ of $q$, while
the rules for updating timers are again referred to
\cite{BrandinWonham:94,Wonham:2011a}.

Let $\Sigma^*$ be the set of all finite strings of elements in
$\Sigma = \Sigma_{act} \dot\cup \{tick\}$, including the empty
string $\epsilon$.  For $\Sigma' \subseteq \Sigma$, the
\emph{natural projection} $P : \Sigma^* \rightarrow \Sigma'^*$ is
defined according to
\begin{equation} \label{eq:natpro}
\begin{split}
P(\epsilon) &= \epsilon, \ \ \epsilon \mbox{ is the empty string;} \\
P(\sigma) &= \left\{
  \begin{array}{ll}
    \epsilon, & \hbox{if $\sigma \notin \Sigma'$,} \\
    \sigma, & \hbox{if $\sigma \in \Sigma'$;}
  \end{array}
\right.\\
P(s\sigma) &= P(s)P(\sigma),\ \ s \in \Sigma^*, \sigma \in \Sigma.
\end{split}
\end{equation}
In the usual way, $P$ is extended to $P : Pwr(\Sigma^*) \rightarrow
Pwr(\Sigma'^*)$, where $Pwr(\cdot)$ denotes powerset. Write $P^{-1}
: Pwr(\Sigma'^*) \rightarrow Pwr(\Sigma^*)$ for the
\emph{inverse-image function} of $P$.

We introduce the languages generated by TDES $\bf G$ in (\ref{e4}). The transition
function $\delta$ is extended to $\delta:Q\times \Sigma^*
\rightarrow Q$ in the usual way. The {\it closed behavior} of $\bf
G$ is the language
\begin{equation} \label{e6}
L({\bf G}) := \{s \in \Sigma^*|\delta(q_0,s)!\}
\end{equation}
and the {\it marked behavior} is
\begin{equation} \label{e7}
L_m({\bf G}) := \{s \in L({\bf G})| \delta(q_0, s) \in Q_m\} \subseteq L({\bf G}).
\end{equation}
We say that $\bf G$ is \emph{nonblocking} if the {\it prefix
closure} (\cite{Wonham:2011a}) $\bar{L}_m({\bf G}) = L({\bf G})$.

To use TDES $\bf G$ in (\ref{e4}) for supervisory control, it is
necessary to specify certain transitions that can be controlled by
an external supervisor. First, as in the untimed theory
\cite{Wonham:2011a}, we need a subset of events that may be
\emph{disabled}. Since disabling an event usually requires
preventing that event indefinitely from occurring, only remote
events belong to this category. Thus let a new subset $\Sigma_{hib}
\subseteq \Sigma_{rem}$ denote the \emph{prohibitible} events; the
supervisor is allowed to disable any prohibitible event. Next, and
specific to TDES, we bring in another category of events which can
\emph{preempt} event $tick$. Note that $tick$ may not be disabled,
inasmuch as no control technology can stop the global clock
indefinitely. On this basis let a new subset $\Sigma_{for} \subseteq
\Sigma_{act}$ denote the {\it forcible} events; a forcible event is
one that preempts event $tick$: if, at a state $q$ of $\bf G$,
$tick$ is defined and so are one or more forcible events, then
$tick$ can be effectively erased from the current list of defined
events (contrast with indefinite erasure)\footnote{One may also
think of forcible events as being able to occur so fast that they
can occur between $ticks$. For a more general use of forcible
events, see \cite{GolaszewskiRamadge87}.}. There is no particular
relation postulated {\it a priori} between $\Sigma_{for}$ and any of
$\Sigma_{hib}$, $\Sigma_{rem}$ or $\Sigma_{spe}$; in particular, a
remote event may be both forcible and prohibitible. It is now
convenient to define the {\it controllable} event set
\begin{equation} \label{e8}
\Sigma_c := \Sigma_{hib}~\dot\cup~\{tick\}.
\end{equation}
Here designating both $\Sigma_{hib}$ and $tick$ controllable is to
simplify terminology. We emphasize that events in $\Sigma_{hib}$ can
be disabled indefinitely, while $tick$ may be preempted only by
events in $\Sigma_{for}$. The {\it uncontrollable} event set
$\Sigma_u$ is
\begin{equation} \label{e8}
\Sigma_{u} := \Sigma - \Sigma_c =  \Sigma_{spe} \dot\cup
(\Sigma_{rem} - \Sigma_{hib}).
\end{equation}

We introduce the notion of controllability as follows. For a string $s \in L({\bf G})$, define
\begin{equation} \label{e11}
Elig_{\bf G}(s):=\{\sigma \in \Sigma|s\sigma \in L({\bf G})\}
\end{equation}
to be the subset of events `eligible' to occur (i.e. defined) at the
state $q = \delta(q_0, s)$. Consider an arbitrary language $F
\subseteq L({\bf G})$ and a string $s \in \overline{F}$; similarly
define the eligible event subset
\begin{equation} \label{e12}
Elig_F(s):= \{\sigma \in \Sigma|s\sigma \in \overline{F}\},
\end{equation}
We say $F$ is {\it controllable} with respect to $\bf G$ in
(\ref{e4}) if, for all $s \in \overline{F}$,
\begin{eqnarray} \label{eq:control}
Elig_F(s)\supseteq
\left\{
   \begin{array}{lcl}
      Elig_{\bf G}(s)\cap(\Sigma_u \dot{\cup}\{tick\}) \\
      ~~~~~~~~~~~~~~~~~\text{if} ~~Elig_F(s)\cap \Sigma_{for} = \emptyset,\\
      Elig_{\bf G}(s)\cap\Sigma_u               \\
      ~~~~~~~~~~~~~~~~~\text{if} ~~Elig_F(s)\cap \Sigma_{for} \neq \emptyset.
   \end{array}
\right.
\end{eqnarray}
Thus $F$ controllable means that an event $\sigma$ is eligible to
occur in $F$ if (i) $\sigma$ is currently eligible in $\bf G$, and
(ii) either $\sigma$ is uncontrollable or $\sigma = tick$ when there
is no forcible event currently eligible in $F$. Recall that in the
untimed supervisory control theory
\cite{RamadgeWonham:87,Wonham:2011a}, $F$ controllable means that
the occurrence of an uncontrollable event in $\bf G$ will not cause
a string $s \in \overline{F}$ to exit from $\overline{F}$; the
difference in TDES is that, the special event $tick$ (formally
controllable) can be preempted only by a forcible event when the
forcible event is eligible to occur.

Whether or not $F$ is controllable, we denote by $\mathcal{C}(F)$
the set of all controllable sublanguages of $F$. Then
$\mathcal{C}(F)$ is nonempty, closed under arbitrary set unions, and
thus contains a unique supremal element denoted by
$sup\mathcal{C}(F)$ \cite{BrandinWonham:94,Wonham:2011a}. Now
consider a specification language $E \subseteq \Sigma^*$ imposed on
the timed behavior of $\bf G$; $E$ may represent a logical and/or
temporal requirement. Let\footnote{${\bf SUP}$ need not be a
(strict) TDES as defined in (\ref{e4}). It can be any automaton
whose event set contains $tick$; we refer to such automata as
\emph{generalized} TDES.}
\begin{equation}\label{e12a}
{\bf SUP} = (X, \Sigma, \xi, x_0, X_m)
\end{equation}
be the corresponding \emph{monolithic supervisor} that is optimal
(i.e., maximally permissive) and nonblocking in the following sense:
$\bf SUP$'s marked language $L_m({\bf SUP})$ is
\begin{equation} \label{e13}
L_m({\bf SUP}) = sup\mathcal{C}(E\cap L_m({\bf G})) \subseteq
L_m(\bf G)
\end{equation}
and moreover its closed language $L({\bf SUP})$ is $L({\bf SUP}) =
\overline{L}_m({\bf SUP}).$
We note that in order to achieve optimal and nonblocking
supervision, $\bf SUP$ should correctly disable prohibitible events
as well as preempt $tick$ via forcible events.


\section{Formulation of Localization Problem}\label{sec:3}

\begin{figure}
\begin{center}
\includegraphics[scale = 0.6]{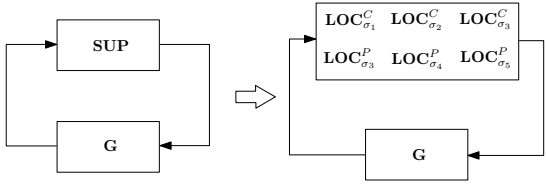}
\caption{Supervisor localization example for illustration: let
$\Sigma_{hib} = \{\sigma_1,\sigma_2,\sigma_3\}$, $\Sigma_{for} =
\{\sigma_3,\sigma_4,\sigma_5\}$; note $\sigma_3 \in \Sigma_{hib}
\cap \Sigma_{for}$.  Localization of $\bf SUP$'s control action
includes two parts: (i) localizing its disabling action into three
\emph{local controllers} $\textbf{LOC}^C_{\sigma_i}$, $i = 1,2,3$,
and (ii) localizing its preemptive action into three \emph{local
preemptors} $\textbf{LOC}^P_{\sigma_j}$, $j = 3,4,5$.}
\label{fig:localization}
\end{center}
\end{figure}

Let TDES $\textbf{G}$ in (\ref{e4}) be the plant to be controlled,
and $E$ be a specification language. Synthesize as in (\ref{e13})
the monolithic optimal and nonblocking supervisor $\bf SUP$;
throughout the paper we assume that $\overline{L}_m({\bf SUP}) \neq
\emptyset$. Supervisor $\bf SUP$'s control action includes (i)
disabling prohibitible events in $\Sigma_{hib}$ and (ii) preempting
$tick$ via forcible events in $\Sigma_{for}$. This section
formulates the localization of $\bf SUP$'s control action with
respect to each prohibitible event as well as to each forcible
event; an illustration of localization is provided in
Fig.~\ref{fig:localization}. Compared to \cite{CaiWonham:2010a}, the
present \emph{supervisor localization} is an extension from untimed
DES to TDES. As will be seen below, the treatment of prohibitible
events is the timed counterpart of the treatment of controllable
events in \cite{CaiWonham:2010a}; on the other hand, localization of
forcible events' preemptive action is specific to TDES,
and we introduce below the new concept ``local preemptor''.
Further, we will discuss applying supervisor localization to the
distributed control of multi-agent TDES.

First, let $\alpha \in \Sigma_{for}$ be an arbitrary forcible event.
We say that ${\bf LOC}^P_{\alpha} = (Y_\alpha, \Sigma_\alpha,
\zeta_\alpha, y_{0, \alpha}, Y_{m, \alpha})$\footnote{${\bf
LOC}^P_{\alpha}$ is a generalized TDES; we further explain this
below in Section~\ref{sec:4}.}, $\Sigma_\alpha \subseteq \Sigma$, is
a \emph{local preemptor} (for $\alpha$) if $\alpha$ is defined at
every state of ${\bf LOC}^P_{\alpha}$ where event $tick$ is
preempted.  Let $P_{\alpha} : \Sigma^* \rightarrow
\Sigma_{\alpha}^*$ be the natural projection as in
(\ref{eq:natpro}).  Then in terms of language, the above condition
means that for every $s \in \Sigma^*$ there holds
\begin{align*} 
&s.tick \in L({\bf G}) ~\&~ s \in P^{-1}_{\alpha} L({\bf
LOC}^P_{\alpha})  ~\&~ \\
&s.tick \notin P^{-1}_{\alpha} L({\bf
LOC}^P_{\alpha})  \Rightarrow  s\alpha \in L({\bf G}) \cap
P^{-1}_{\alpha} L({\bf LOC}^P_{\alpha}).
\end{align*}
Notation $s.tick$ means that event $tick$ occurs after string $s$,
and will be used henceforth. The left side of the above implication
means that event $tick$ is preempted in ${\bf LOC}^P_{\alpha}$ after
string $s$ (after $s$ event $tick$ is defined in $L({\bf G})$ but
not in ${\bf LOC}^P_{\alpha}$), and the right side says that
forcible event $\alpha$ is defined in ${\bf LOC}^P_{\alpha}$ (and in
$L({\bf G})$) after $s$.  That is, forcible event $\alpha$ acts to
preempt $tick$. The event set $\Sigma_\alpha$ of ${\bf
LOC}^P_{\alpha}$ in general satisfies $\{\alpha, tick\}\subseteq
\Sigma_\alpha \subseteq \Sigma$; in typical cases, however, both
subset containments are strict, as will be illustrated in
Section~\ref{sec:5:exmp}. Also, for simplicity we assume the lower
and upper time bounds of events in $\Sigma_\alpha$ coincide with the
bounds on the corresponding events in $\Sigma$ (this is, in fact,
guaranteed by the localization procedure presented below in
Section~\ref{sec:4}). It is worth emphasizing that $\Sigma_\alpha$
(precisely defined below) is not fixed {\it a priori}, but will be
systematically determined, as part of our localization result, to
ensure correct preemptive action.

Next, let $\beta \in \Sigma_{hib}$ be an arbitrary prohibitible
event. We say that ${\bf LOC}^C_{\beta} = (Y_\beta, \Sigma_\beta,
\zeta_\beta, y_{0,\beta}, Y_{m,\beta})$, $\Sigma_\beta \subseteq
\Sigma$, is a {\it local controller} (for $\beta$) if ${\bf
LOC}^C_{\beta}$ can disable only event $\beta$. Let
$P_{\beta} : \Sigma^* \rightarrow \Sigma_{\beta}^*$ be the natural
projection as in (\ref{eq:natpro}). Then in terms of language, the
above condition means that for all $s \in \Sigma^*$ and $\sigma \in
\Sigma$, there holds (cf. \cite{CaiWonham:2010a})
\begin{align*}
s\sigma \in L({\bf G}) ~\&~ &s\in P^{-1}_{\alpha} L({\bf
LOC}^C_{\beta}) ~\&~ \\
 &s\sigma \notin P^{-1}_{\alpha} L({\bf
LOC}^C_{\beta})  \Rightarrow  \sigma = \beta.
\end{align*}
The event set $\Sigma_\beta$ of ${\bf LOC}^C_{\beta}$ in general
satisfies $\{\beta\}\subseteq \Sigma_\beta \subseteq
\Sigma$.\footnote{Event set $\Sigma_\beta$ need not contain event
$tick$, since ${\bf LOC}^C_{\beta}$'s disabling action may be purely
logical and irrelevant to time.\label{fnote:beta}} Like
$\Sigma_\alpha$ above, $\Sigma_\beta$ will be generated as part of
our localization result to guarantee correct disabling action;
again, the events in $\Sigma_\beta$ are assumed to have the same
lower and upper time bounds as the corresponding events in $\Sigma$.

Now we formulate the {\it Supervisor Localization Problem} of TDES:
Construct a set of local preemptors  $\{{\bf LOC}^P_{\alpha}|\alpha
\in \Sigma_{for}\}$ and a set of local controllers $\{{\bf
LOC}^C_{\beta}|\beta \in \Sigma_{hib}\}$, with
\begin{align}
   L({\bf LOC}) := \left(\mathop \bigcap\limits_{\alpha \in \Sigma_{for}}P_\alpha^{-1}L({\bf LOC}^P_{\alpha}) \right) \notag\\
   \cap \left(\mathop \bigcap\limits_{\beta \in \Sigma_{hib}}P_\beta^{-1}L({\bf LOC}^C_{\beta}) \right) \label{e20}\\
   L_m({\bf LOC}) := \left(\mathop \bigcap\limits_{\alpha \in \Sigma_{for}}P_\alpha^{-1}L_m({\bf LOC}^P_{\alpha})\right)\notag \\
   \cap \left(\mathop \bigcap\limits_{\beta \in \Sigma_{hib}}P_\beta^{-1}L_m({\bf LOC}^C_{\beta}) \right) \label{e21}
\end{align}
such that {\bf LOC} is \emph{control equivalent} to {\bf SUP} (with
respect to {\bf G}) in the following sense:
\begin{align*}
   L({\bf G}) \cap L({\bf LOC}) &= L({\bf SUP}), \\
   L_m({\bf G}) \cap L_m({\bf LOC}) &= L_m({\bf SUP}).
\end{align*}

For the sake of easy implementation and comprehensibility, it would
be desired in practice that the state sizes of local
preemptors/controllers be very much less than that of their parent
monolithic supervisor. Inasmuch as this property is neither precise
to state nor always achievable, it is omitted from the above formal
problem statement; in applications, nevertheless, it should be kept
in mind.

\begin{figure}[!t]
\begin{center}
\includegraphics[scale = 0.6]{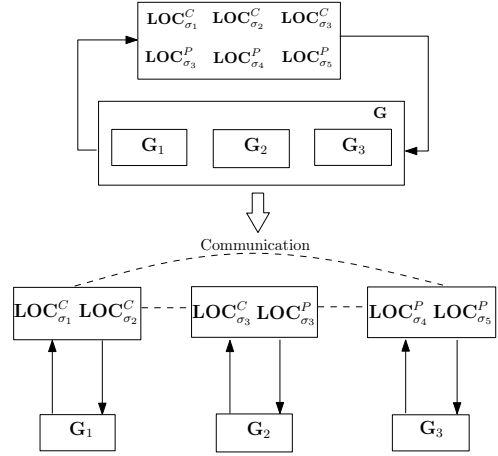}
\caption{Example of distributed control by allocating local
preemptors/controllers. Continuing the example in
Fig.~\ref{fig:localization}, let plant $\textbf{G}$ be composed of
three agents ${\bf G}_k$ with event sets $\Sigma_k$, $k \in [1,3]$.
Suppose $\sigma_1,\sigma_2 \in \Sigma_1$, $\sigma_2, \sigma_3 \in
\Sigma_2$, and $\sigma_3, \sigma_4, \sigma_5 \in \Sigma_3$; thus
${\bf G}_1$ and ${\bf G}_2$ share event $\sigma_2$, and ${\bf G}_2$
and ${\bf G}_3$ share event $\sigma_3$. Then a convenient allocation
is displayed, where each local controller/preemptor is owned by
exactly one agent. The allocation creates a distributed control
architecture for the multi-agent plant, in which each agent acts
semi-autonomously while interacting with other agents through
communication of shared events.} \label{fig:allocation}
\end{center}
\end{figure}

Using a set of local preemptors and local controllers that is
control equivalent to {\bf SUP}, we can build an optimal and
nonblocking \emph{distributed control architecture} for a
multi-agent TDES plant. Let the plant $\textbf{G}$ with event set
$\Sigma$ be composed\footnote{Composition of multiple TDES involves
first taking \emph{synchronous product} of the untimed DES, and then
unifying the time bounds of shared events
\cite{BrandinWonham:94,Wonham:2011a}.} of $n$ component TDES (or
agents) ${\bf G}_k$ with $\Sigma_k$ ($k \in
[1,n]$).\footnote{ Note that each ${\bf G}_k$ may contain
multiple forcible and/or prohibitible events.} According to
(\ref{e4}), $\Sigma_k = \Sigma_{act,k} \dot{\cup} \{tick\}$ (event
$tick$ is shared by all agents); thus $\Sigma = \bigcup_{k=1}^n
\Sigma_{act,k} \dot{\cup} \{tick\}$. In addition to $tick$, we also
allow the $\Sigma_{act,k}$ to share events. Now let $\Sigma_{for,k},
\Sigma_{hib,k} \subseteq \Sigma_k$ be the forcible event set and
prohibitible event set, respectively, of agent ${\bf G}_k$; then
$\Sigma_{for} = \bigcup_{k=1}^n \Sigma_{for,k}$ and $\Sigma_{hib} =
\bigcup_{k=1}^n \Sigma_{hib,k}$. For each forcible event $\alpha \in
\Sigma_{for}$ there is a local preemptor ${\bf LOC}^P_{\alpha}$; and
for each prohibitible event $\beta \in \Sigma_{hib}$ there is a
local controller ${\bf LOC}^C_{\beta}$. { These local
preemptors/controllers need to be allocated among individual agents,
for each agent may have multiple forcible/prohibitible events.} A
convenient allocation is to let each local controller/preemptor be
owned by exactly one agent; an example is displayed in
Fig~\ref{fig:allocation}. Choosing this or (obvious) alternative
ways of allocation would be case-dependent.

\section{Procedure of Supervisor Localization} \label{sec:4}

We solve the Supervisor Localization Problem of TDES by developing a
localization procedure for the supervisor's preemptive and disabling
action, respectively. The procedure extends the untimed counterpart
in \cite{CaiWonham:2010a}. In particular, localizing the
supervisor's preemption of event $tick$ with respect to each
individual forcible event is novel in the current TDES setup,
for which we introduce below two new ideas ``preemption
consistency relation'' and ``preemption cover''.

Given a TDES plant ${\bf G} = (Q, \Sigma, \delta, q_0, Q_m)$ (as in
(\ref{e4})) and a corresponding monolithic supervisor ${\bf SUP} =
(X, \Sigma, \xi, x_0, X_m)$ (as in (\ref{e12a})) with respect to an
imposed specification, we present the localization of {\bf SUP}'s
preemptive and disabling action in the sequel.

\subsection{Localization of Preemptive Action} \label{sec:4.1}

Fix an arbitrary forcible event $\alpha \in \Sigma_{for}$.
First define $E_{tick}: X \rightarrow \{1,0\}$ according to
\begin{equation}\label{e23}
E_{tick}(x) = 1~\text{iff}~\xi(x,tick)!.
\end{equation}
Thus $E_{tick}(x) = 1$ means that $tick$ is defined at
state $x$ in $\bf SUP$. Next define $F_\alpha:X \rightarrow \{1,0\}$
according to $F_\alpha(x) = 1$ iff
\begin{align} \label{e24}
\xi(x,\alpha)! ~\&~ \neg\xi(x,tick)! ~\&~  (\exists s\in
\Sigma^*) \notag\\
\Big( \xi(x_0,s)=x ~\&~\delta(q_0, s.tick)! \Big)
\end{align}
So $F_\alpha(x) = 1$ means that forcible event $\alpha$ is defined
at state $x$ (i.e. $\xi(x,\alpha)!$), which effectively preempts the
occurrence of event $tick$ (i.e. $tick$ is not defined at $x$ in
$\bf SUP$ but is defined at some state in the plant $\textbf{G}$
corresponding to $x$ via string $s$). It should be noted that at
state $x$, $\alpha$ need not be the only forcible event that
preempts $tick$, for there can be other forcible events, say
$\alpha'$, defined at $x$.  In that case, by (\ref{e24})
$F_{\alpha'}(x) = 1$ as well.

Based on the preemption information captured by $E_{tick}$ and
$F_{\alpha}$ above, we define the following binary relation
$\mathcal{R}^P_{\alpha}$ (for $\alpha$) on $X$, called `preemption
consistency'. { This relation determines if two states of
{\bf SUP} have consistent preemptive action with respect to the
forcible event $\alpha$.}

\begin{defn} \label{def1}
Let $\mathcal{R}^P_{\alpha} \subseteq X \times X$. We say that
$\mathcal{R}^P_{\alpha}$ is a {\it preemption consistency relation}
with respect to $\alpha \in \Sigma_{for}$ if for every $x,x' \in X$,
$(x,x')\in \mathcal{R}^P_{\alpha}$ iff \begin{equation} \label{e25}
E_{tick}(x)\cdot F_\alpha(x') = 0 = E_{tick}(x')\cdot F_\alpha(x).
\end{equation}
\end{defn}

\begin{figure}[!t]
    \begin{minipage}{0.45\linewidth}\centering
    \end{minipage}\hfill
    \begin{minipage}{0.5\linewidth}
        \includegraphics[scale = 0.48]{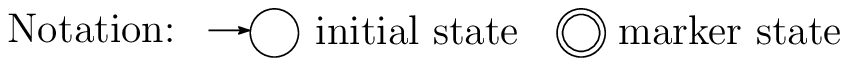}\\
    \end{minipage}\\
    \begin{minipage}{0.45\linewidth}\centering
      $\alpha \in \Sigma_{for}$
    \end{minipage}\hfill
    \begin{minipage}{0.5\linewidth}
        \begin{overpic}[scale = 0.45]{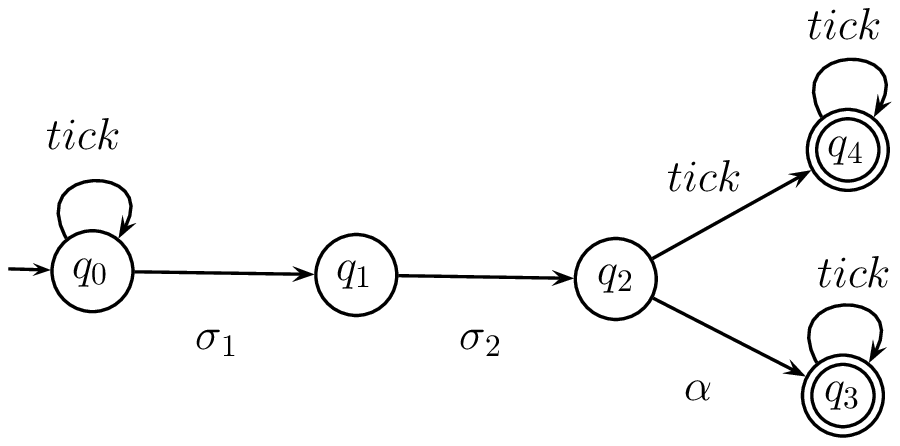}
            \put(50, 0){$\bf G$}
        \end{overpic}
    \end{minipage}\\
    \begin{minipage}{0.45\linewidth}\centering
    \includegraphics[scale = 0.8]{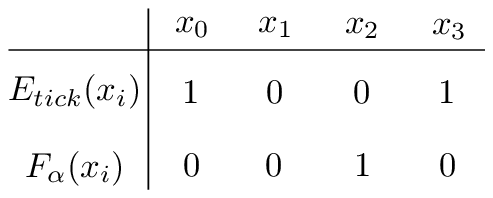}
    \end{minipage}
    \hfill
    \begin{minipage}{0.5\linewidth}
    \vspace{2em}
        \begin{overpic}[scale = 0.45]{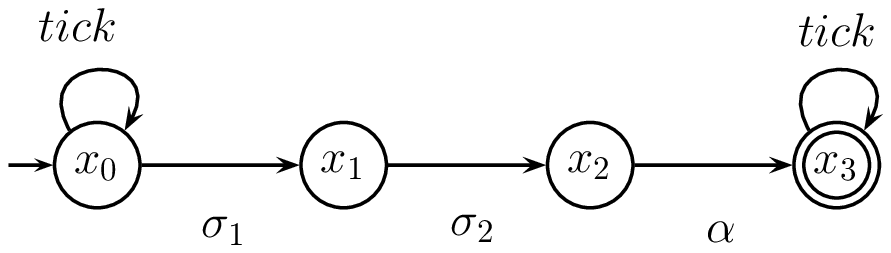}
            \put(45,-10){$\bf SUP$}
        \end{overpic}
    \end{minipage}
    \vspace{1em}
\caption{Preemption consistency relation is not transitive: $(x_0,
x_1) \in \mathcal{R}^P_{\alpha}$, $(x_1, x_2) \in
\mathcal{R}^P_{\alpha}$, but $(x_0, x_2) \notin
\mathcal{R}^P_{\alpha}$.} \label{fig:transitive}
\end{figure}

Thus a pair of states $(x,x')$ in ${\bf SUP}$ is \emph{not}
preemption consistent with respect to $\alpha$ only when $tick$ is
defined at $x$ but is preempted by $\alpha$ at $x'$, or vice versa.
Otherwise, $x$ and $x'$ are preemption consistent, i.e. $(x,x') \in
\mathcal{R}^P_{\alpha}$. It is easily verified that
$\mathcal{R}^P_{\alpha}$ is reflexive and symmetric, but not
transitive; an illustration is provided in Fig.~\ref{fig:transitive}. Hence
$\mathcal{R}^P_{\alpha}$ is not an equivalence relation. This fact
leads to the following definition of a {\it preemption cover}.
Recall that a {\it cover} on a set $X$ is a family of nonempty
subsets (or {\it cells}) of $X$ whose union is $X$.

\begin{defn} \label{def2}
Let $I$ be some index set, and $\mathcal{C}^P_{\alpha} = \{X_i\subseteq X|i \in I\}$ a cover on $X$. We say that $\mathcal{C}^P_{\alpha}$ is a {\it preemption cover} with respect to $\alpha$ if
\begin{align} \label{e26}
(\rmnum{1})~&(\forall i\in I, \forall x, x' \in X_i) (x,x') \in \mathcal{R}^P_{\alpha}, \notag\\
(\rmnum{2})~&(\forall i\in I, \forall \sigma \in \Sigma)\Big[(\exists x\in X_i)\xi(x,\sigma)! \Rightarrow \\
&\big((\exists j\in I)(\forall x'\in X_i)\xi(x',\sigma)!\Rightarrow
\xi(x',\sigma)\in X_j\big)\Big]. \notag
\end{align}
\end{defn}

A preemption cover $\mathcal{C}^P_{\alpha}$ lumps states of $\bf
SUP$ into (possibly overlapping) cells $X_i$, $i\in I$. According to
(\rmnum{1}) all states that reside in a cell $X_i$ must be pairwise
preemption consistent; and (\rmnum{2}) for every event $\sigma\in
\Sigma$, all states that can be reached from any states in $X_i$ by
a one-step transition $\sigma$ must be covered by the same cell
$X_j$. Inductively, two states $x,x'$ belong to a common cell of
$\mathcal{C}^P_{\alpha}$ if and only if { $x$ and $x'$
are preemption consistent, and two future states, say $y$ and $y'$,
that can be reached respectively from $x$ and $x'$ by a given string
are again preemption consistent.} We say that a preemption cover
$\mathcal{C}^P_{\alpha}$ is a {\it preemption congruence} if
$\mathcal{C}^P_{\alpha}$ happens to be a {\it partition} on $X$,
namely its cells are pairwise disjoint.

Having defined a preemption cover $\mathcal{C}^P_{\alpha}$ on $X$,
we construct, below, a local preemptor ${\bf LOC}^P_{\alpha} =
(Y_\alpha, \Sigma_\alpha, \zeta_\alpha, y_{0,\alpha},$
$Y_{m,\alpha})$ for the forcible event $\alpha$ to preempt $tick$.

({\bf Step 1}) The state set is $Y_\alpha := I$, with each state $y
\in Y_\alpha$ being a cell $X_i$ of the cover
$\mathcal{C}^P_{\alpha}$. In particular, the initial state
$y_{0,\alpha}$ is a cell $X_{i0}$ where $x_0$ belongs, i.e. $x_0 \in
X_{i0}$, and the marker state set $Y_{m,\alpha}:=\{i\in I|X_i\cap
X_m \neq \emptyset\}$.

({\bf Step 2}) For the event set $\Sigma_\alpha$, define the
transition function $\zeta_\alpha':I\times\Sigma \rightarrow I$ over
the entire event set $\Sigma$ by $\zeta_\alpha'(i,\sigma)=j$ if
\begin{align} \label{e277}
(\exists x\in X_i)&\xi(x,\sigma)\in X_j \mbox{\ \ \ and \ \ \ }\notag\\
&(\forall x'\in X_i)\big[\xi(x',\sigma)! \Rightarrow
\xi(x',\sigma)\in X_j\big].
\end{align}
Choose $\Sigma_\alpha$ to be the union of $\{\alpha, tick\}$ with
other events which are {\it not} selfloop transitions of
$\zeta_\alpha'$, i.e.
\begin{align}\label{e27}
\Sigma_\alpha :=  \{\alpha, tick\}\dot\cup \{\sigma \in \Sigma -
\{\alpha, tick\} \ | \ (\exists i,j \in I) \notag\\
i \neq j \ \& \
\zeta_\alpha'(i,\sigma) = j\}.
\end{align}
Intuitively, only those non-selfloop transitions may
affect decisions on \emph{tick} preemption, and thus the events that
are only selfloops may be removed. Note that $\{\alpha,
tick\}\subseteq\Sigma_\alpha \subseteq \Sigma$.

({\bf Step 3}) Define the transition function $\zeta_\alpha$ to be
the restriction of $\zeta_\alpha'$ to $\Sigma_\alpha$; namely
$\zeta_\alpha := \zeta_\alpha'|_{\Sigma_\alpha}: I \times
\Sigma_\alpha \rightarrow I$ { according to
$\zeta_\alpha(i,\sigma) = \zeta'_\alpha(i,\sigma)$ for every $i \in
I$ and $\sigma \in \Sigma_\alpha$.}

We note that ${\bf LOC}^P_{\alpha}$ thus constructed is not a TDES
as defined in (\ref{e4}), for its states do not contain timer
information. ${\bf LOC}^P_{\alpha}$ is indeed a generalized TDES
because its event set $\Sigma_\alpha$ contains $tick$. We will be
concerned only with its behavior, namely its closed and marked
languages. Also note that, owing to possible overlapping of cells in
the cover $\mathcal{C}^P_{\alpha}$, the choices of $y_{0,\alpha}$
and $\zeta_{\alpha}$ may not be unique, and consequently ${\bf
LOC}^P_{\alpha}$ may not be unique. In that case we pick an
arbitrary instance of ${\bf LOC}^P_{\alpha}$. If
$\mathcal{C}^P_{\alpha}$ happens to be a preemption congruence,
however, then ${\bf LOC}^P_{\alpha}$ is unique.

By the same procedure, we generate a set of local preemptors ${\bf
LOC}^P_{\alpha}$, one for each forcible event $\alpha \in
\Sigma_{for}$.  We will verify below that these generated preemptors
collectively achieve the same preemptive action of event $tick$ as
the monolithic supervisor $\bf SUP$ does.

\subsection{Localization of Disabling Action} \label{sec:4.2}

Next, we turn to the localization of ${\bf SUP}$'s disabling action,
which is analogous to the treatment in \cite{CaiWonham:2010a}. Fix
an arbitrary prohibitible event $\beta \in \Sigma_{hib}$. First
define $E_\beta: X \rightarrow \{1,0\}$ according to
\begin{equation}\label{e28}
E_\beta(x) = 1~\text{iff}~\xi(x,\beta)!
\end{equation}
So $E_\beta(x) = 1$ means that $\beta$ is defined at state $x$ in
$\bf SUP$. Next define $D_\beta: X\rightarrow\{1,0\}$ according to
$D_\beta(x) = 1$ iff
\begin{eqnarray}\label{e29}
\neg\xi(x,\beta)! \ ~\&~\ (\exists s\in \Sigma^*)\left(\xi(x_0,s)=x
~\&~\delta(q_0, s\beta)! \right)
\end{eqnarray}
Thus $D_\beta(x) = 1$ means that $\beta$ must be disabled at $x$
(i.e. $\beta$ is disabled at $x$ in $\bf SUP$ but is defined at some
state in the plant $\textbf{G}$ corresponding to $x$ via string
$s$). In addition, define $M: X \rightarrow \{1,0\}$ according to
\begin{equation} \label{e33}
M(x) = 1~\text{iff}~ x \in X_m.
\end{equation}
Thus $M(x)=1$ means that state $x$ is marked in $\bf SUP$. Finally
define $T:X\rightarrow \{1,0\}$ according to
\begin{align} \label{e34}
T(x) = 1~\text{iff}~ (\exists s\in \Sigma^*)\xi(x_0,s) = x  \&
\delta(q_0,s) \in Q_m
\end{align}
So $T(x)=1$ means that some state, corresponding to
$x$ via $s$, is marked in $\bf G$. Note that for each $x\in X$, it
follows from $L_m({\bf SUP}) \subseteq L_m({\bf G})$ that $T(x) = 0
\Rightarrow M(x) = 0$ and $M(x) = 1 \Rightarrow T(x) = 1$
\cite{CaiWonham:2010a}.

Based on (\ref{e28})-(\ref{e34}), we define the following binary
relation $\mathcal{R}^C_{\beta} \subseteq X\times X$, called {\it
control consistency} with respect to prohibitible event $\beta$ (cf.
\cite{CaiWonham:2010a}), according to $(x,x')\in
\mathcal{R}^C_{\beta}$ iff
\begin{align} \label{e30}
&(\rmnum{1})~E_\beta(x)\cdot D_\beta(x') = 0 = E_\beta(x')\cdot D_\beta(x), \notag\\
&(\rmnum{2})~T(x) = T(x') \Rightarrow M(x) = M(x').
\end{align}

Thus a pair of states $(x,x')$ in $\bf SUP$ satisfies $(x,x') \in
\mathcal{R}^C_{\beta}$ if (\rmnum{1}) { event $\beta$ is
defined at one state, but not disabled at the other}; and
(\rmnum{2}) $x$ and $x'$ are both marked or both unmarked in $\bf
SUP$, provided both are marked or unmarked in $\bf G$. It is easily
verified that $\mathcal{R}^C_{\beta}$ is generally not transitive
\cite{CaiWonham:2010a}, thus not an equivalence relation. Now let
$I$ be some index set, and $\mathcal{C}^C_{\beta} = \{X_i\subseteq
X|i \in I\}$ a cover on $X$. Similar to Definition~\ref{def2}, we
define $\mathcal{C}^C_{\beta}$ to be a {\it control cover} with
respect to $\beta$ if
\begin{align} \label{e31}
(\rmnum{1})~&(\forall i\in I, \forall x, x' \in X_i) (x,x') \in \mathcal{R}^C_{\beta}, \notag\\
(\rmnum{2})~&(\forall i\in I, \forall \sigma \in \Sigma)\Big[(\exists x\in X_i)\xi(x,\sigma)! \Rightarrow \\
&\big((\exists j\in I)(\forall x'\in X_i)\xi(x',\sigma)! \Rightarrow
\xi(x',\sigma)\in X_j\big)\Big]. \notag
\end{align}
{ Note that the only difference between control cover and
preemption cover in Definition \ref{def2} is the binary relation
(control consistency $\mathcal{R}^C_{\beta}$ or preemption
consistency $\mathcal{R}^P_{\alpha}$) used in condition
(\rmnum{1}).}

With the control cover $\mathcal{C}^C_{\beta}$ on $X$, we construct
by the same steps ({\bf Step1}) - ({\bf Step 3}), above, a local
controller ${\bf LOC}^C_{\beta} = (Y_\beta, \Sigma_\beta,
\zeta_\beta, y_{0,\beta}, Y_{m,\beta})$ for prohibitible event
$\beta$. Here, the choice of event set $\Sigma_\beta$ is
{ (cf. (\ref{e27}))}
\begin{align}\label{e32}
\Sigma_\beta :=  \{\beta\}\dot\cup \{\sigma \in \Sigma - \{\beta\} \
|\ (\exists i,j \in I)\notag \\
i \neq j \ \&\ \zeta_\beta'(i,\sigma) = j\}.
\end{align}

$\Sigma_\beta$ need not contain event $tick$, as noted in Footnote~\ref{fnote:beta}.
As before, owing to possible overlapping of cells in the control
cover $\mathcal{C}^C_{\beta}$, a local controller ${\bf
LOC}^C_{\beta}$ need not be unique. If, however,
$\mathcal{C}^C_{\beta}$ happens to be a control congruence (i.e.
$\mathcal{C}^C_{\beta}$ is a partition on $X$), then ${\bf
LOC}^C_{\beta}$ is unique. In the same way, we generate a set of
local controllers ${\bf LOC}^C_{\beta}$, one for each prohibitible
event $\beta \in \Sigma_{hib}$. We will verify that the collective
disabling action of these local controllers is identical to that of
the monolithic supervisor $\bf SUP$.

Finally, notice that an event $\beta$ may be both prohibitible and
forcible. In that case, $\beta$ will be equipped with both a local
controller which exercises disabling action specific to $\beta$, and
a local preemptor which implements preemption of event $tick$ via
$\beta$. It appears that here a conflict could arise: $\beta$'s
local preemptor intends to use $\beta$ to preempt $tick$, but
$\beta$ is disabled by its local controller.  However, since
$\beta$'s local preemptor and controller are both derived from $\bf
SUP$ which is proved to contain no such conflict
\cite{BrandinWonham:94,Wonham:2011a}, the conflict indeed cannot
arise between $\beta$'s local preemptor and controller. Our main
result below confirms this fact.

\subsection{Main Result}\label{sec:4.3:main}

Here is the main result of this section, which states that the local
preemptors and controllers generated by the proposed localization
procedure collectively achieve the monolithic optimal and
nonblocking supervision.

\begin{thm} \label{thm1}
The set of local preemptors $\{{\bf LOC}^P_{\alpha}|\alpha \in
\Sigma_{for}\}$ and the set of local controllers $\{{\bf
LOC}^C_{\beta}|\beta \in \Sigma_{hib}\}$ constructed above solve the
Supervisor Localization Problem; that is,
\begin{align}
   L({\bf G}) \cap L({\bf LOC}) &= L({\bf SUP}), \label{e38}\\
   L_m({\bf G}) \cap L_m({\bf LOC}) &= L_m({\bf SUP}). \label{e39}
\end{align}
where $L({\bf LOC})$ and $L_m({\bf LOC})$ are as defined in (\ref{e20}) and (\ref{e21}), respectively.
\end{thm}

Theorem~\ref{thm1} extends the untimed supervisor localization
result in \cite{CaiWonham:2010a} to the TDES setup, where not only
the disabling action but also the $tick$-preemptive action of the
monolithic supervisor needs to be localized.  Thus supervisor
localization in TDES generates a set of local controllers,
 one for each individual prohibitible event, as well as
a set of local preemptors, one for each individual
forcible event. The proof of Theorem~\ref{thm1}, below,
relies on the concepts of TDES controllability, control cover, as
well as preemption cover.

{ Since for every preemption cover (resp. control cover),
the presented procedure constructs a local preemptor (resp.
preemption cover),} Theorem~\ref{thm1} asserts that every set of
preemption and control covers together generates a solution to the
Supervisor Localization Problem. In particular, a set of {\it
state-minimal} local preemptors (resp. local controllers), possibly
non-unique, can in principle be defined from a set of suitable
preemption covers (resp. control covers). The minimal state problem,
however, is known to be NP-hard \cite{SuWonham:2004}. In
\cite{CaiWonham:2010a} we proposed, nevertheless, a polynomial-time
localization algorithm which computes congruences instead of covers;
and empirical evidence was given that significant state size
reduction can often be achieved. That localization algorithm (see
\cite[Section~III-B]{CaiWonham:2010a}) for untimed DES can easily be
adapted in the current TDES case, the only modification being to use
the new definitions of preemption and control consistency given in
Sections~\ref{sec:4.1} and \ref{sec:4.2}.

So far we have focused on localization of the
monolithic supervisor. In fact, the developed localization procedure
may be applied to decompose a modular (decentralized or
hierarchical) supervisor just as well.  Thus when a TDES is
large-scale and the monolithic supervisor not feasibly computable,
we may in principle combine localization with an effective modular
supervisory synthesis: first compute a set of modular supervisors
which achieves the same behavior as the monolithic supervisor, and
then apply localization to decompose each modular supervisor in the
set. This is done in \cite{CaiWonham:2010a,CaiWonham:2010b} for
large-scale untimed DES; and we aim to work out the timed
counterpart in future research.


We now provide the proof of Theorem~\ref{thm1}. Equation (\ref{e39})
and the $(\supseteq)$ direction of (\ref{e38}) may be verified
analogously as in \cite{CaiWonham:2010a}. For completeness we
present the verification in the Appendix. Here we prove
$(\subseteq)$ in (\ref{e38}), which involves the TDES's
controllability definition, preemption consistency, and control
consistency.

\emph{Proof of Theorem~\ref{thm1}.} ($\subseteq$, \ref{e38}) We show
this by induction. First, the empty string $\epsilon$ belongs to
$L({\bf G})$, $L({\bf LOC})$, and $L({\bf SUP})$, because these
languages are all nonempty.  Next, suppose $s \in L({\bf G}) \cap
L({\bf LOC})$, $s \in L({\bf SUP})$, and $s\sigma \in L({\bf G})
\cap L({\bf LOC})$ for an arbitrary event $\sigma \in \Sigma$. It
will be proved that $s\sigma \in L({\bf SUP})$. Since $\Sigma =
\Sigma_u \ \dot\cup\ \Sigma_c = \Sigma_u \ \dot\cup\ \{tick\} \
\dot\cup\ \Sigma_{hib}$ (as in (\ref{e8})), we consider the
following three cases.

(\rmnum{1}) Let $\sigma \in \Sigma_u$. Since $L_m(\bf SUP)$ is
controllable (see (\ref{eq:control})), and $s\sigma \in L({\bf G})$
(i.e. $\sigma \in Elig_{\bf G}(s)$ by (\ref{e11})), we have $\sigma
\in Elig_{L_m({\bf SUP})}(s)$. That is, by (\ref{e12}) $s\sigma \in
\overline{L}_m({\bf SUP}) = L({\bf SUP})$.

\begin{figure}[!t]
\centering
    \includegraphics[scale = 0.7]{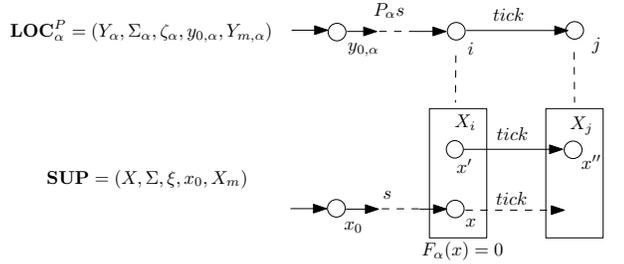}
\caption{State-transition correspondences between ${\bf
LOC}^P_\alpha$ and $\bf SUP$.  It is proved in the text that $tick$
is also defined at state $x$.} \label{fig:theorem}
\end{figure}

(\rmnum{2}) Let $\sigma = tick$. We will show $tick \in
Elig_{L_m({\bf SUP})}(s)$ to conclude that $s.tick \in
\overline{L}_m({\bf SUP}) = L({\bf SUP})$. By the hypothesis that
$s$, $s.tick \in L({\bf LOC})$ and equation (\ref{e20}), for every
forcible event $\alpha \in \Sigma_{for}$ there holds $s, s.tick \in
P_\alpha^{-1}L({\bf LOC}^P_{\alpha})$, i.e. $P_\alpha(s),
P_\alpha(s)\ tick \in L({\bf LOC}^P_{\alpha})$. Recall ${\bf
LOC}^P_{\alpha} = (Y_\alpha, \Sigma_\alpha, \zeta_\alpha,
y_{0,\alpha}, Y_{m,\alpha})$, and let $i :=
\zeta_\alpha(y_{0,\alpha}, P_\alpha(s))$ and $j := \zeta_\alpha(i,
tick)$. By definition of $\zeta_\alpha'$ in (\ref{e277}), any
$\sigma \notin \Sigma_\alpha$ (defined in (\ref{e27})) is only a
selfloop transition of $\zeta_\alpha'$; hence
$\zeta_\alpha'(y_{0,\alpha}, s)=i$.  By (\ref{e277}) again, there
exist $x, x' \in X_i$ and $x'' \in X_j$ such that $\xi(x_0,s)=x$ and
$\xi(x', tick) = x''$ in {\bf SUP}. These state-transition
correspondences between ${\bf LOC}^P_{\alpha}$ and {\bf SUP} are
displayed in Fig.~\ref{fig:theorem}.

Now that $x,x'$ belong to the same cell $X_i$, by the preemption
cover definition (Definition~\ref{def2}) $x$ and $x'$ must be
preemption consistent, i.e. $(x, x') \in \mathcal {R}^P_{\alpha}$.
Since $\xi(x', tick)!$, by (\ref{e23}) we have $E_{tick}(x') = 1$.
Thus the requirement $E_{tick}(x') \cdot F_\alpha(x) = 0$
(Definition~\ref{def1}) yields that $F_\alpha(x) = 0$. The latter,
by (\ref{e24}), gives rise to the following three cases: (Case 1)
$\neg \xi(x,\alpha)!$, (Case 2) $\xi(x,tick)!$, or (Case 3) $(\neg
\exists s\in \Sigma^*)\big( \xi(x_0,s)=x ~\&~\delta(q_0,
s.tick)!\big)$. First, Case 3 is impossible, because by the
hypothesis that $s \in L({\bf SUP})$ and $s.tick \in L({\bf G})$ we
have $\xi(x_0,s)!$ and $\delta(q_0, s.tick)!$.  Next, Case 2 means
directly $tick \in Elig_{L_m({\bf SUP})}(s)$. Finally, Case 1
implies { $\alpha \notin Elig_{L_m({\bf SUP})}(s)$}; note
that this holds for all $\alpha \in \Sigma_{for}$.  Hence
$Elig_{L_m({\bf SUP})}(s) \cap \Sigma_{for} = \emptyset$.  Then by
the fact that ${\bf SUP}$ is controllable, we derive from
(\ref{eq:control}) that $tick \in Elig_{L_m({\bf SUP})}(s)$.

(\rmnum{3}) Let $\sigma \in \Sigma_{hib}$. By the hypothesis $s$,
$s\sigma \in L({\bf LOC})$ and equation (\ref{e20}), we have
$s,s\sigma \in P_\sigma^{-1}L({\bf LOC}^C_{\sigma})$, i.e.
$P_\sigma(s), P_\sigma(s)\sigma \in L({\bf LOC}^C_{\sigma})$. As in
(\rmnum{2}), let
$i:=\zeta_\sigma(y_{0,\sigma},P_\sigma(s))=\zeta'_\sigma(y_{0,\sigma},s)$
and $j := \zeta_\sigma(i,\sigma)$. By the definition of
$\zeta_\sigma'$ in (\ref{e277}), there exist $x, x' \in X_i$,
$x''\in X_j$ such that $\xi(x_0,s)=x$ and $\xi(x',\sigma) = x''$.
Since $x,x'$ belong to the same cell $X_i$, by the control cover
definition $x$ and $x'$ must be control consistent, i.e. $(x, x')
\in \mathcal {R}^C_{\sigma}$. That $\xi(x', \sigma)!$ implies by
(\ref{e28}) that $E_{\sigma}(x') = 1$. Thus the requirement
$E_{\sigma}(x') \cdot D_\sigma(x) = 0$ yields that $D_\sigma(x) =
0$.   The latter, by (\ref{e29}), gives rise to the following two
cases: (Case 1) $\xi(x,\sigma)!$, or (Case 2) $(\neg \exists s\in
\Sigma^*)\xi(x_0,s)=x ~\&~\delta(q_0, s\sigma)!$. Case 2 is
impossible, because by the hypothesis that $s \in L({\bf SUP})$ and
$s.tick \in L({\bf G})$ we have $\xi(x_0,s)!$ and $\delta(q_0,
s.tick)!$. But in Case 1, $\xi(x,\sigma)!$ i.e. $s\sigma \in L({\bf
SUP})$. \hfill $\square$

\section{Case Study: Manufacturing Cell}\label{sec:5:exmp}

\begin{figure}[!t]
\centering
    \begin{minipage}{0.5\linewidth}
    \centering
        \includegraphics[scale = 0.5]{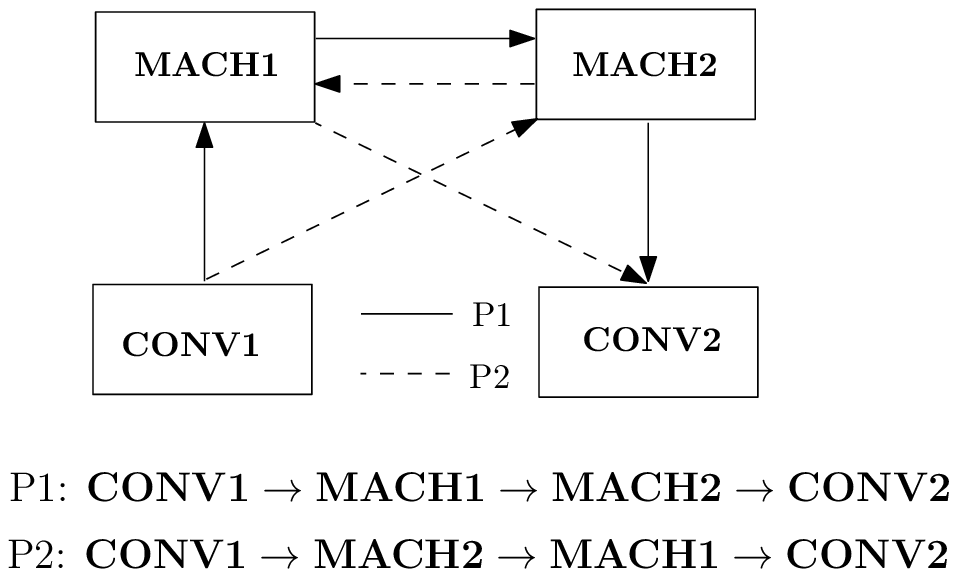}
    \end{minipage}
    \hfill
    \begin{minipage}{0.45\linewidth}
    \centering
        \begin{overpic}[scale = 0.45]{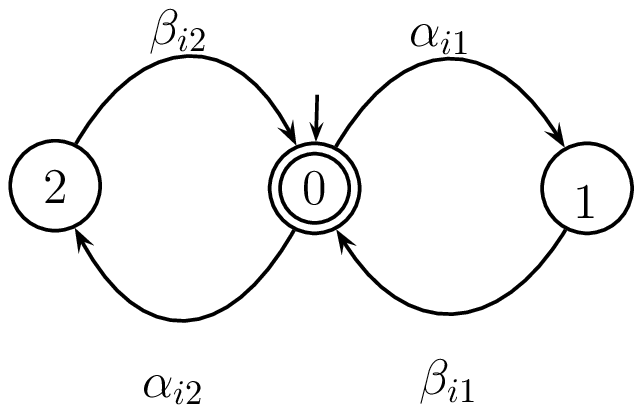}
            \put(-5,-10){\scriptsize Untimed DES models of}
            \put(10,-20){\scriptsize $\bf MACHi$, i = 1, 2}
        \end{overpic}\\
    \end{minipage}
\caption{Manufacturing Cell} \label{fig1}
\end{figure}

We illustrate supervisor localization in TDES by studying a
manufacturing cell example, taken from
\cite{BrandinWonham:94},\cite[Section~9.11]{Wonham:2011a}. As
displayed in Fig.~\ref{fig1}, the cell consists of two machines,
$\bf MACH1$ and $\bf MACH2$, an input conveyor $\bf CONV1$ as an
infinite source of workpieces, and output conveyor $\bf CONV2$ as an
infinite sink. Each machine processes two types of parts, P1 and P2.
Each type of part is routed as shown in Fig.~\ref{fig1}.  The
untimed DES models of the machines are also displayed in
Fig.~\ref{fig1}; here $\alpha_{ij}$ ($i,j\in[1,2]$) is the event
``$\bf MACHi$ starts to work on a Pj-part'', while $\beta_{ij}$
($i,j\in[1,2]$) is ``$\bf MACHi$ finishes working on a Pj-part''.
Assign lower and upper time bounds to each event, with the notation
(event, lower bound, upper bound), as follows:
\begin{align*}
{\bf MACH1}&\text{'s timed events}: \\
&(\alpha_{11},1,\infty)~~(\beta_{11},3,3)~~(\alpha_{12},1,\infty)~~(\beta_{12},2,2)\\
{\bf MACH2}&\text{'s timed events}: \\
&(\alpha_{21},1,\infty)~~(\beta_{21},1,1)~~(\alpha_{22},1,\infty)~~(\beta_{22},4,4)
\end{align*}
So $\alpha_{ij}$ are remote events (upper bound $\infty$), and
$\beta_{ij}$ prospective events (finite upper bounds).  Now
the TDES models of the two machines can be generated \cite[p.425]{Wonham:2011a}. Their joint behavior is the synchronous product of
the two TDES, which in this example is the plant to be controlled.
\begin{figure}[!t]
\centering
    \begin{minipage}{0.4\linewidth}\vspace{2em}
    \centering
        \begin{overpic}[scale = 0.4]{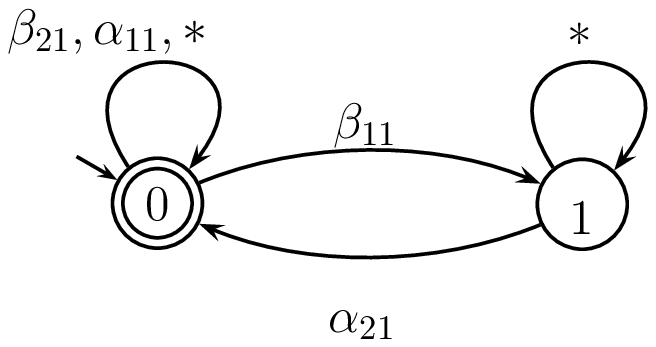}
            \put(0,-10){\small $* = \{tick, \alpha_{12},\alpha_{22},$}
            \put(40,-25){\small $\beta_{12},\beta_{22}\}$}
            \put(40,-40){\small $\bf SPEC1$}
        \end{overpic}
    \end{minipage}
    \hfill
    \begin{minipage}{0.55\linewidth}
    \centering
        \begin{overpic}[scale = 0.45]{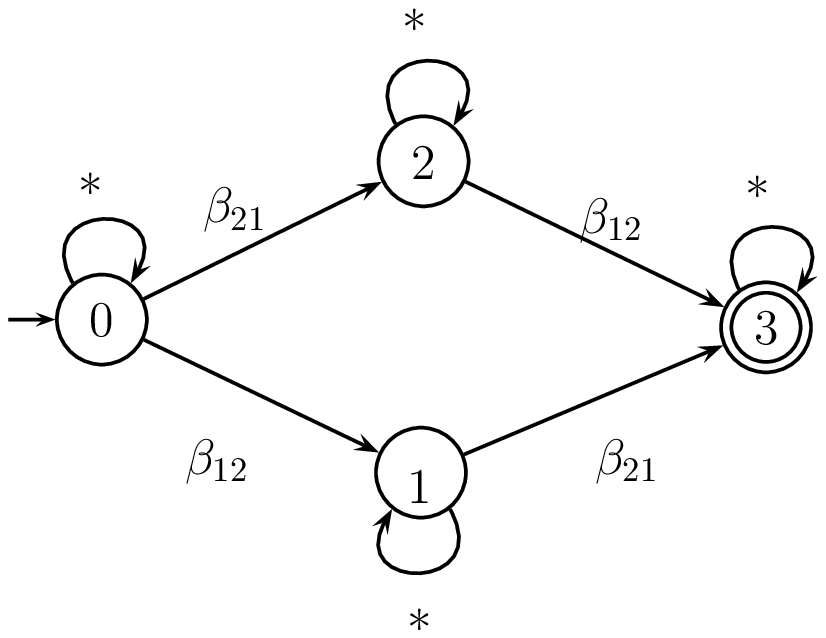}
            \put(2,-10){\small $* = \{tick, \alpha_{11},\alpha_{12},\alpha_{21},\alpha_{22},$}
            \put(40, -20){\small $\beta_{11},\beta_{22}\}$}
            \put(40,-30){\small $\bf SPEC3$}
        \end{overpic}
    \end{minipage}\\
    \vspace{1em}
    \begin{minipage}{0.4\linewidth}
    \centering\vspace{3em}
        \begin{overpic}[scale = 0.4]{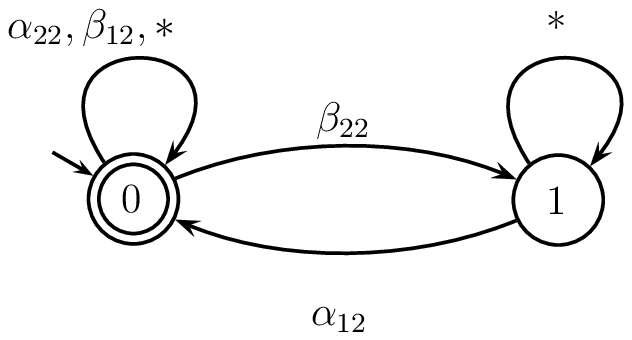}
            \put(5,-10){\small $* = \{tick, \alpha_{11},\alpha_{21},$}
            \put(40, -25){\small $\beta_{11},\beta_{21}\}$}
            \put(40,-40){\small $\bf SPEC2$}
        \end{overpic}
    \end{minipage}
    \hfill
    \begin{minipage}{0.55\linewidth}\vspace{3em}
    \centering
        \begin{overpic}[scale = 0.5]{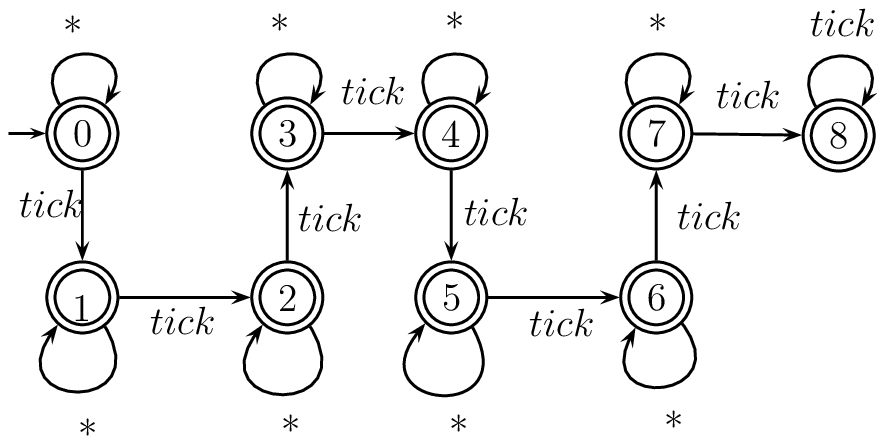}
            \put(15,-10){\small $* = \{\alpha_{11},\alpha_{12},\alpha_{21},\alpha_{22}, $}
            \put(30, -20){\small $\beta_{11},\beta_{12},\beta_{21},\beta_{22}\}$}
            \put(40,-30){\small $\bf SPEC4$}
        \end{overpic}
    \end{minipage}
\vspace{3.5em} \caption{Control specifications: logical and
temporal.  The marked state 3 of $\bf SPEC3$ corresponds
to the completion of a production cycle: one P1-part and one P2-part
are processed by both machines.}
 \label{fig2}
\end{figure}

To impose behavioral constraints on the two machines' joint
behavior, we take the events $\alpha_{ij}$ to be both prohibitible and forcible,
i.e. $\Sigma_{hib}=\Sigma_{for} = \{\alpha_{ij}|i,j =
1,2\}$, and the $\beta_{ij}$ to be uncontrollable, i.e. $\Sigma_u
= \{\beta_{ij}|i,j = 1,2\}$. We impose the following {\it logical}
control specifications as well as a {\it temporal} specification:

(S1) A P1-part must be processed first by {\bf MACH1} and then by
{\bf MACH2}.

(S2) A P2-part must be processed first by {\bf MACH2} and then by
{\bf MACH1}.

(S3) One P1-part and one P2-part must be processed in a production
cycle.

(S4) A production cycle must be completed in at most 8 time
units.\footnote{Here we choose ``8 time units'' because it is,
according to \cite{BrandinWonham:94, Wonham:2011a}, the minimal time
to complete one production cycle. Thus this temporal specification
represents a time-minimization requirement.}

\noindent These four specifications are formalized as automata $\bf
SPEC1$, $\bf SPEC2$, $\bf SPEC3$, and $\bf SPEC4$, respectively, as
displayed in Fig.~\ref{fig2}.  The temporal specification $\bf
SPEC4$ is simply an $8$-$tick$ sequence, with all states marked;
$\bf SPEC4$ forces any TDES with which it is synchronized to halt
after at most 8 $ticks$, i.e. after 8 $ticks$ to execute no further
event whatever except event $tick$. Thus it extracts the marked
strings (if any) which satisfy this constraint, namely the `tasks'
of TDES that can be accomplished in at most 8 $ticks$ (which turns
out to be exactly one production cycle according to
\cite{BrandinWonham:94, Wonham:2011a}).

\begin{figure}[!t]
\centering
    \includegraphics[scale = 0.7]{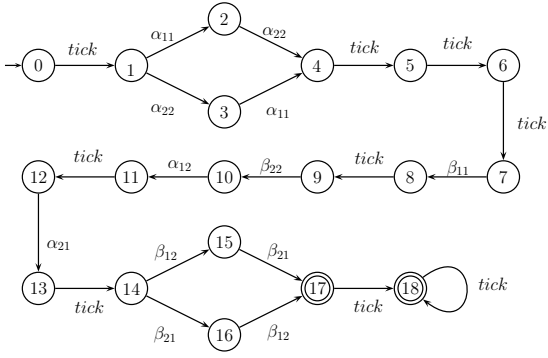}
\caption{Monolithic optimal and nonblocking supervisor {\bf SUP}.}
\label{fig:sup}
\end{figure}

Now the plant to be controlled is the synchronous product of TDES
{\bf MACH1} and {\bf MACH2} \cite[p.425]{Wonham:2011a}, and the overall
control specification is the synchronous product of automata $\bf
SPEC1$-$\bf SPEC4$ in Fig.~\ref{fig2}.  We compute as in (\ref{e13})
the corresponding monolithic optimal and nonblocking supervisor {\bf
SUP}; the computation is done by the $supcon$ command in XPTTCT
\cite{Wonham:2008}. {\bf SUP} has 19 states and 21 transitions, as
displayed in Fig.~\ref{fig:sup}. We see that {\bf SUP} represents
the behavior that the manufacturing cell accomplishes exactly one
working cycle, within $8$ ticks, producing one P1-part and one
P2-part. Indeed, each event is executed exactly once, and each
forcible event preempts $tick$ immediately after it becomes eligible
to occur.

We now apply supervisor localization to decompose the monolithic
supervisor {\bf SUP} into local preemptors and local controllers,
respectively for each forcible event and each prohibitible event.
Specifically, since $\Sigma_{hib}=\Sigma_{for} = \{\alpha_{ij}|i,j =
1,2\}$, we will compute a local preemptor and a local controller for
each $\alpha_{ij}$, responsible for $\alpha_{ij}$'s
$tick$-preemptive action and its disabling action, respectively.
This computation can be done by an algorithm adapted from
\cite{CaiWonham:2010a} (as discussed in Section~\ref{sec:4.3:main});
here, however, owing to the simple (chain-like) structure of {\bf
SUP} (Fig.~\ref{fig:sup}), local preemptors/controllers can be
derived by inspection. We demonstrate such a derivation below, which
results in a local preemptor ${\bf LOC}^P_{\alpha_{11}}$ for the
forcible (and prohibitible) event $\alpha_{11}$.  Other derivations
of local preemptors/controllers are similar.

To derive a local preemptor ${\bf LOC}^P_{\alpha_{11}}$ for event
$\alpha_{11}$, we find a preemption cover $\mathcal
{C}^P_{\alpha_{11}}$ for $\alpha_{11}$ on $\bf SUP$'s state set as
follows.  Initialize $\mathcal {C}^P_{\alpha_{11}}$ to be $\mathcal
{C}^P_{\alpha_{11}} = \big\{[0], [1], [2], ..., [18]\big\}$, i.e.
each cell contains exactly one state of $\bf SUP$.  Subsequently,
we merge as many cells together as possible according to
Definitions~\ref{def1} and \ref{def2}, while maintaining $\mathcal
{C}^P_{\alpha_{11}}$ to be a preemption cover.

(\rmnum{1}) Cells $[0]$ and $[1]$ cannot be merged. Since
$E_{tick}(0)=1$ (event $tick$ is defined at state $0$) and
$F_{\alpha_{11}}(1)=1$ ($tick$ is preempted by $\alpha_{11}$ at
state $1$), the pair of states $(0,1)$ is not preemption consistent,
i.e. $(0,1) \notin \mathcal{R}^P_{\alpha_{11}}$. Consequently,
merging cells $[0]$ and $[1]$ violates requirement (i) of preemption
cover (Definition~\ref{def2}).

(\rmnum{2}) Cells $[1], [3]$ and cells $[2], [4]$ can be merged. For
cells $[2]$ and $[4]$, we have $F_{\alpha_{11}}(2)=0$, $E_{tick}(2) = 0$ ($tick$ is
preempted at state $2$, but by $\alpha_{22}$ not by $\alpha_{11}$)
and $E_{tick}(4)=1$, $F_{\alpha_{11}}(4) = 0$ (event $tick$ is defined at state $4$). Thus
$(2,4) \in \mathcal{R}^P_{\alpha_{11}}$, which satisfies requirement
(i) of preemption cover.  Moreover since no common event is defined
on states $2$ and $4$, requirement (ii) of preemption cover is
trivially satisfied. Therefore cells $[2], [4]$ can be merged.

For cells $[1]$ and $[3]$, we have
$F_{\alpha_{11}}(1)=F_{\alpha_{11}}(3)=1$ ($tick$ is preempted by
$\alpha_{11}$ at both states $1$ and $3$) and
$E_{tick}(1)=E_{tick}(3)=0$.  Thus $(1,3) \in
\mathcal{R}^P_{\alpha_{11}}$, which satisfies requirement (i) of
preemption cover. Now event $\alpha_{11}$ is defined at both states
$1$ and $3$, but it leads to states $2$ and $4$ respectively, which
have been verified to be preemption consistent.  Hence, requirement
(ii) of preemption cover is also satisfied, and cells $[1], [3]$ can
be merged.

By merging the above two pairs of cells, we derive $\mathcal
{C}^P_{\alpha_{11}}= \big\{[0], [1,3], [2, 4], [5], ...,
[18]\big\}$.

(\rmnum{3}) Cells $[2, 4], [5], \ldots, [18]$ can all be merged
together.  Note, indeed, that $F_{\alpha_{11}}(\cdot)=0$ for all
these states (no $tick$ preemption by $\alpha_{11}$). On checking
the preemption consistency and preemption cover definitions as
above, we conclude that the final preemption cover is $\mathcal
{C}^P_{\alpha_{11}} = \big\{[0], [1,3], [2, 4, 5, ..., 18]\big\}$.
It is in fact a preemption congruence.

\begin{figure}[!t]
\centering
    \begin{minipage}{0.45\linewidth}
    \centering
        \begin{overpic}[scale = 0.5]{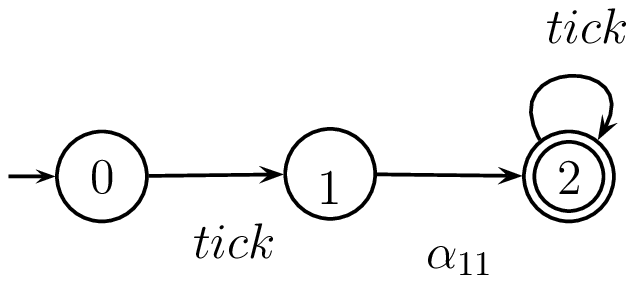}
            \put(35,-12){${\bf LOC}^P_{\alpha_{11}}$}
        \end{overpic}
    \end{minipage}
    \hfill
    \begin{minipage}{0.45\linewidth}
    \centering
        \begin{overpic}[scale = 0.5]{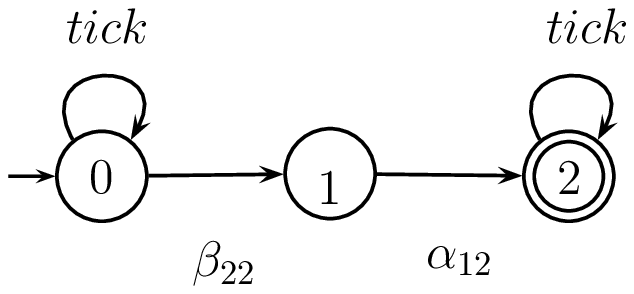}
            \put(35,-12){${\bf LOC}^P_{\alpha_{12}}$}
        \end{overpic}
    \end{minipage}\\
    \vspace{2em}
    \begin{minipage}{0.45\linewidth}
    \centering
        \begin{overpic}[scale = 0.5]{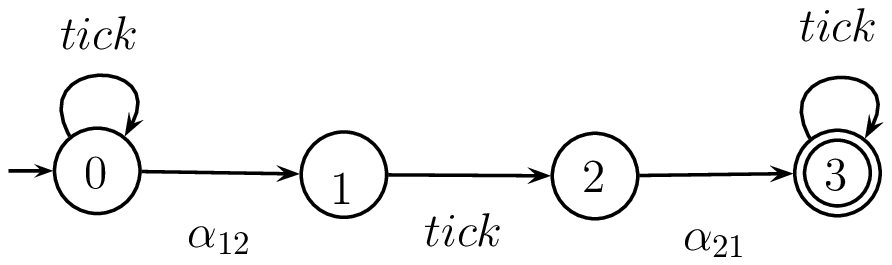}
            \put(40,-12){${\bf LOC}^P_{\alpha_{21}}$}
        \end{overpic}
    \end{minipage}
    \hfill
    \begin{minipage}{0.45\linewidth}
    \centering
        \begin{overpic}[scale = 0.5]{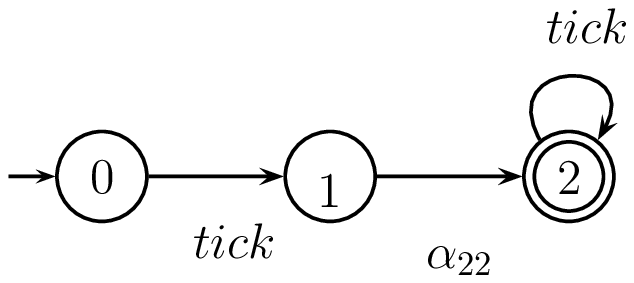}
            \put(35,-12){${\bf LOC}^P_{\alpha_{22}}$}
        \end{overpic}
    \end{minipage}\\
\vspace{1em} \caption{Local preemptors for individual forcible
events. The alphabet of each local preemptor is the set
of events displayed in each automaton.} \label{fig3}
\end{figure}

\begin{figure}[!t]
\centering
    \begin{minipage}{0.45\linewidth}
    \centering
        \begin{overpic}[scale = 0.5]{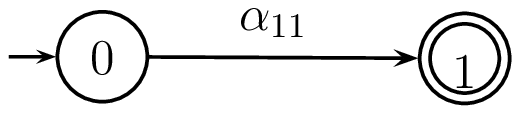}
            \put(40,-12){${\bf LOC}^C_{\alpha_{11}}$}
        \end{overpic}
    \end{minipage}
    \begin{minipage}{0.45\linewidth}
    \centering
        \begin{overpic}[scale = 0.5]{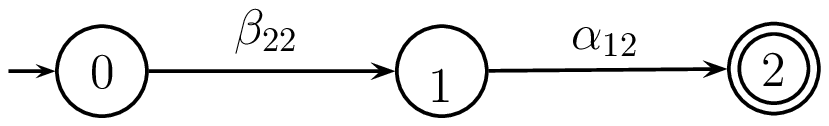}
            \put(42,-12){${\bf LOC}^C_{\alpha_{12}}$}
        \end{overpic}
    \end{minipage}\\
    \begin{minipage}{0.45\linewidth}\vspace{3em}
    \centering
        \begin{overpic}[scale = 0.5]{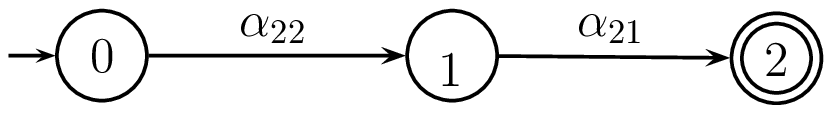}
            \put(40,-12){${\bf LOC}^C_{\alpha_{21}}$}
        \end{overpic}
    \end{minipage}
    \begin{minipage}{0.45\linewidth}\vspace{3em}
    \centering
        \begin{overpic}[scale = 0.5]{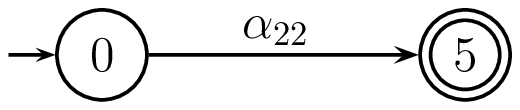}
            \put(40,-12){${\bf LOC}^C_{\alpha_{22}}$}
        \end{overpic}
    \end{minipage}\\
\vspace{1em} \caption{Local controllers for individual prohibitible
events.  The alphabet of each local controller is the set
of events displayed in each automaton.} \label{fig4}
\end{figure}

Having found the preemption cover $\mathcal {C}^P_{\alpha_{11}}$, we
apply \textbf{(Step 1) - (Step 3)} in Section~\ref{sec:4.1} to
construct a local preemptor ${\bf LOC}^P_{\alpha_{11}}$, with
transition structure displayed in Fig.~\ref{fig3}. Note that the
event set of ${\bf LOC}^P_{\alpha_{11}}$ is exactly $\{\alpha_{11},
tick\}$, which means that ${\bf LOC}^P_{\alpha_{11}}$ does not need
to observe any external events in order to execute its preemptive action.
Similarly, we derive other local preemptors and local
controllers, all displayed in Figs.~\ref{fig3} and \ref{fig4}. Here,
for example, the event set of ${\bf LOC}^P_{\alpha_{12}}$ is
$\{\alpha_{12}, tick, \beta_{22}\}$; so event $\beta_{22}$ originating in
$\bf MACH2$ has to be observed by ${\bf LOC}^P_{\alpha_{12}}$. We
have then verified that their joint behavior (via synchronous
product) is identical to the monolithic optimal and nonblocking
behavior of {\bf SUP}, i.e. (\ref{e38}) and (\ref{e39}) hold.

We see that each local preemptor/controller has fewer states, with a
simpler structure, than the monolithic {\bf SUP}; this renders each
one's preemptive/disabling action more transparent.  For example,
the local preemptor ${\bf LOC}^P_{\alpha_{11}}$ (resp. ${\bf
LOC}^P_{\alpha_{22}}$) in Fig.~\ref{fig3} means that after one tick,
forcible event $\alpha_{11}$ preempts event $tick$ and {\bf MACH1}
starts to work on a P1-part (resp. $\alpha_{22}$ preempts $tick$ and
{\bf MACH2} works on a P2-part).  This is possible because
$\alpha_{11}$ (resp. $\alpha_{22}$) has lower time bound $1$ and
becomes eligible to occur after one tick. For another example, the
local preemptor ${\bf LOC}^P_{\alpha_{21}}$ in Fig.~\ref{fig3}
specifies that after occurrence of $\alpha_{12}$ followed by a
$tick$, forcible event $\alpha_{21}$ preempts $tick$ and {\bf MACH2}
starts to work on a P1-part. This preemption is due to the fact that
$\alpha_{21}$ has lower time bound $1$ and becomes eligible to occur
after occurrence of $\beta_{22}$ plus one tick
(according to Fig.~\ref{fig:sup} event $\alpha_{22}$ first occurs in
$\bf MACH2$, which implies from the untimed model in Fig.~\ref{fig1}
the event order $\alpha_{22}.\beta_{22}.\alpha_{21}$). But
occurrence of $\alpha_{12}$ implies that $\beta_{22}$ has just
occurred (see Fig.~\ref{fig:sup}).

For control logic, the local controller ${\bf LOC}^C_{\alpha_{12}}$
in Fig.~\ref{fig4} means that prohibitible event $\alpha_{12}$ is
enabled only after occurrence of event $\beta_{22}$, i.e. {\bf
MACH1} starts to work on a P2-part only after {\bf MACH2} finishes
that P2-part. On the other hand, the logic of ${\bf
LOC}^C_{\alpha_{21}}$ is a bit subtle; it specifies that
prohibitible event $\alpha_{21}$ is enabled after occurrence of
event $\alpha_{22}$.  At first glance, the logic seems to violate
the specification $\bf SPEC1$ in Fig.~\ref{fig2}, which says that
$\alpha_{21}$ should not be enabled before occurrence of
$\beta_{11}$. Observe, nevertheless, that $\alpha_{21}$ cannot
become eligible to occur before occurrence of $\beta_{22}$ which has
lower (and upper) time bound $4$, and event $\beta_{11}$ in fact has
already occurred when $\beta_{22}$ occurs (see Fig.~\ref{fig:sup}).
Hence it is legal to enable $\alpha_{21}$ after $\alpha_{22}$.

\begin{figure}[!t]
\centering
    \includegraphics[scale = 0.7]{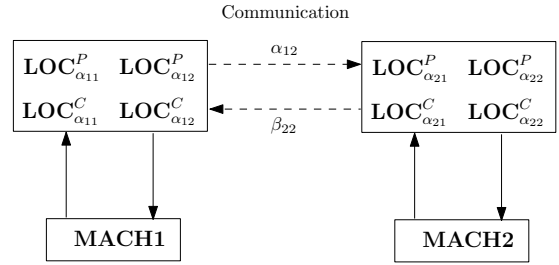}
\caption{Distributed control architecture for manufacturing cell.}
\label{fig:distribution}
\end{figure}

Finally, with the derived set of local preemptors and controllers,
we build a distributed control architecture for this manufacturing
cell of two machines; see Fig.~\ref{fig:distribution}.  Each machine
acquires those local preemptors/controllers with respect to its own
distinct forcible/prohibitible events, thereby being capable of
executing local preemptive/disabling actions.  For these local
actions to jointly achieve the same controlled behavior as the
monolithic supervisor does, communicating certain `critical' events
(in this case $\alpha_{12}$ and $\beta_{22}$) between the two
machines is essential. { The critical events are obtained
by intersecting the alphabet of one machine and the alphabets of
local preemptors/controllers of the other machine.}

\section{Conclusions} \label{sec:conclusion}

We have established supervisor localization in the Brandin-Wonham
timed DES framework. Under this localization scheme, each individual
agent disables its own prohibitible events and preempts event $tick$
via its own forcible events; overall, these local control actions
collectively achieve monolithic optimal and nonblocking supervision.
We have demonstrated the timed supervisor localization on a
manufacturing cell case study.  In future research, we
aim to combine the developed localization approach with an effective
modular supervisor synthesis to address distributed control of
large-scale real-time DES.

\bibliographystyle{plain}
\bibliography{zhang_references}

\appendix
\section{Appendix}

We complete the proof of Theorem~\ref{thm1}, namely
equation (\ref{e39}) and $(\supseteq)$ in (\ref{e38}).

($\supseteq$, \ref{e39}) Since $L_m({\bf SUP}) \subseteq L_m({\bf
G})$, it suffices to show that $L_m({\bf SUP}) \subseteq L_m({\bf
LOC})$. That is, by (\ref{e21}),
\begin{align} \label{eq:app1}
&(\forall \alpha \in \Sigma_{for})\ L_m({\bf SUP}) \subseteq
P_\alpha^{-1}L_m({\bf LOC}^P_{\alpha}),\\
\label{eq:app2}&(\forall \beta \in \Sigma_{hib})L_m({\bf SUP})
\subseteq P_\beta^{-1}L_m({\bf LOC}^C_{\beta}).
\end{align}
We prove (\ref{eq:app1}), and (\ref{eq:app2}) follows similarly. Let
$s = \sigma_0\sigma_1\cdots\sigma_h \in L_m({\bf SUP})$. Then
$x_1:=\xi(x_0,\sigma_0)$, ..., $x_{h+1} := \xi(x_0,s) \in X_m$. By
the construction of ${\bf LOC}^P_{\alpha}$ ($\alpha \in
\Sigma_{for}$ arbitrary), in particular the transition function
$\zeta_\alpha'$ over $\Sigma$ in (\ref{e277}), there exist $i_0,
i_1, ..., i_{h+1}$ with $(i_0 = y_{0,\alpha})$ such that
\begin{align} \label{e40}
x_0 \in X_{i_0}~&\&~ \zeta_\alpha'(i_0,\sigma_0) = i_1, \notag \\
x_1 \in X_{i_1}~&\&~ \zeta_\alpha'(i_1,\sigma_1) = i_2, \notag\\
&\vdots \\
x_{h+1} \in X_{i_{h+1}}~&\&~ \zeta_\alpha'(i_h,\sigma_h) = i_{h+1}. \notag
\end{align}
So $\zeta_\alpha'(i_0,\sigma_0\sigma_1\cdots\sigma_h)
=\zeta_\alpha'(i_0,s)!$, and belongs to $Y_{m,\alpha}$ because
$X_{i_{h+1}} \cap X_m \neq \emptyset$ ($x_{h+1}$ belongs to $X_m$).
Moreover since any $\sigma \notin \Sigma_\alpha$ (defined in
(\ref{e27})) is only a selfloop transition of $\zeta_\alpha'$, we
derive $\zeta_\alpha(i_0,P_\alpha(s)) \in Y_{m,\alpha}$. Hence,
$P_\alpha(s) \in L_m({\bf LOC}^P_{\alpha})$, i.e. $s\in
P_\alpha^{-1}L_m({\bf LOC}^P_{\alpha})$.

($\supseteq$, \ref{e38}) This is an easy consequence of ($\supseteq$, \ref{e39}):
\begin{align*}
L({\bf SUP}) &= \overline{L_m({\bf SUP})} \\
             &\subseteq \overline{L_m({\bf G}) \cap L_m({\bf LOC})} \\
             &\subseteq L({\bf G}) \cap L({\bf LOC}).
\end{align*}

($\subseteq$, \ref{e39}) Let $s \in L_m({\bf G}) \cap L_m({\bf
LOC})$; by (\ref{e21}), for every $\beta \in \Sigma_{hib}$, $s \in
P_\beta^{-1}L_m({\bf LOC}^C_{\beta})$, i.e. $P_\beta(s) \in L_m({\bf
LOC}^C_{\beta})$. Write $i:= \zeta_\beta(y_{0,\beta},P_\beta(s))$.
Then there exists $x \in X_{i} \cap X_m$; thus $M(x) = 1$ (defined
in (\ref{e33})), which also implies $T(x) = 1$ (defined in
(\ref{e34})). On the other hand, since $L_m({\bf G}) \cap L_m({\bf
LOC}) \subseteq L({\bf G}) \cap L({\bf LOC}) = L({\bf SUP})$ (the
last equality has already been shown), we have $s \in L({\bf SUP})$.
That is, $\xi(x_0,s)!$; as in (\ref{e40}) above we derive
$\xi(x_0,s) \in X_{i}$, and by the control cover definition
(\ref{e31}) it holds that $(x,\xi(x_0,s)) \in \mathcal
{R}^C_{\beta}$. Since $s \in L_m({\bf G})$, i.e. $\delta(q_0,s) \in
Q_m$, we have $T(\xi(x_0, s)) = 1$. Therefore by requirement (ii) of
the control consistency definition (\ref{e30}), $M(\xi(x_0, s)) =
1$, i.e. $s \in L_m({\bf SUP})$.

\end{document}